\documentclass[12pt]{iopart}
\usepackage{hyperref}
\usepackage{listings}
\usepackage{graphicx}
\expandafter\let\csname equation*\endcsname\relax
\expandafter\let\csname endequation*\endcsname\relax
\usepackage{amsmath}
\usepackage[sorting=none]{biblatex}
\bibliography{papers}
\usepackage{enumitem}
\usepackage{booktabs}
\usepackage[separate-uncertainty=true]{siunitx}
\usepackage{verbatim}
\usepackage{tcolorbox}
\usepackage{newfloat}
\DeclareFloatingEnvironment[fileext=lop]{example}

\renewcommand{\theenumi}{\arabic{enumi}}
\DeclareUnicodeCharacter{03B1}{$\alpha$}
\DeclareUnicodeCharacter{03B2}{$\beta$}
\DeclareUnicodeCharacter{03B3}{$\gamma$}

\usepackage[margin=1in, bottom=1in]{geometry}

\usepackage{fancyhdr}
\begin{document}
\title{Evaluating AI and Human Authorship Quality in Academic Writing through Physics Essays}
\author{Will Yeadon, Elise Agra, Oto-obong Inyang, Paul Mackay, Arin Mizouri}
\address{Department of Physics, Durham University, Lower Mountjoy, South Rd, Durham, DH1 3LE, UK}
\ead{will.yeadon@durham.ac.uk}
\vspace{10pt}
\begin{indented}
\item[]{March 2024}
\end{indented}

\begin{abstract}
This study evaluates $n = 300$ short-form physics essay submissions, equally divided between student work submitted before the introduction of ChatGPT and those generated by OpenAI's GPT-4. In blinded evaluations conducted by five independent markers who were unaware of the origin of the essays, we observed no statistically significant differences in scores between essays authored by humans and those produced by AI (p-value $= 0.107$, $\alpha$ = 0.05). Additionally, when the markers subsequently attempted to identify the authorship of the essays on a 4-point Likert scale - from `Definitely AI' to `Definitely Human' - their performance was only marginally better than random chance. This outcome not only underscores the convergence of AI and human authorship quality but also highlights the difficulty of discerning AI-generated content solely through human judgment. Furthermore, the effectiveness of five commercially available software tools for identifying essay authorship was evaluated. Among these, ZeroGPT was the most accurate, achieving a 98\% accuracy rate and a precision score of 1.0 when its classifications were reduced to binary outcomes. This result is a source of potential optimism for maintaining assessment integrity. Finally, we propose that texts with $\leq 50\%$ AI-generated content should be considered the upper limit for classification as human-authored, a boundary inclusive of a future with ubiquitous AI assistance whilst also respecting human-authorship.
\end{abstract}

\section{Introduction}\label{intro}
\subsection{Background}
\label{subsection-bkgd}
The year 2023 marked a pivotal moment in the integration of AI text-completion technologies within educational settings worldwide. Educators were confronted with the reality that students could use tools like ChatGPT to instantly complete assignments, sparking fears of academic dishonesty and the undermining of the educational process. In response to these challenges, some institutions ceased assigning traditional homework, opting instead for in-depth preparatory work that fosters discussion and assessment within the classroom \cite{timesChatGPT}. 

The proficiency of modern Large Language Models (LLMs) in generating text across a broad spectrum of topics, from science \cite{scienceChatGPT} to finance \cite{financeChatGPT}, is well-documented. The advancements in these models are evident when comparing their current outputs to those from five years ago, showcasing a significant improvement in quality. This progression has prompted academics to investigate the capabilities of AI in composing essays on a wide range of subjects. Notably, studies have demonstrated that AI-written documents excel in tasks ranging from the analysis of general legal principles \cite{lawChatGPT} to Old English Poetry, including intricate analyses of works like Beowulf \cite{beowolf}.

In response to the rapid advancements in AI and their potential impact on academic integrity, this study investigates the quality of AI-authored compared to human-authored essays in an essay writing task for an accredited physics course at Durham University, UK. Employing a blinded assessment methodology, markers evaluated essays without knowledge of their origin to eliminate bias and focus on content quality and adherence to academic standards. Essays were selected from submissions prior to the widespread adoption of modern LLMs, ensuring a fair comparison by minimizing potential AI-generated content in the control group. Authorship was assigned post-evaluation to maintain the integrity of the assessment process. This approach allows for a critical examination of AI's capabilities in producing academically sound essays and an investigation into the broader implications for educational practices and assessment integrity. As LLMs evolve, understanding their benefits and limitations is crucial for educators, students, and academic institutions.

\subsection{State of the Art}
\label{sota}
There are many powerful LLMs available, and benchmarking them is a comprehensive field of study aimed at evaluating the capabilities of these models across a broad spectrum of cognitive and linguistic tasks \cite{systematicChatGPT}. As of this writing, OpenAI's GPT-4 consistently ranks at or near the top of most LLM benchmark leaderboards, scoring highly on prominent benchmarks such as the Massive Multitask Language Understanding (MMLU), according to its technical report \cite{gpt-4-technical-report}.

The MMLU benchmark, known for its extensive range of subjects from humanities to hard sciences, provides a rigorous test of a model's ability to understand and generate responses across diverse knowledge areas. Similarly, the SuperGLUE benchmark \cite{superglue}, which focuses on tasks that require a deep understanding of language, such as question answering, inference, and reasoning, challenges models to demonstrate advanced levels of comprehension and the ability to handle nuanced linguistic constructs. GPT-4's performance, independently supported by scoring over 90\% on SuperGLUE \cite{superglue-independent}, underscores its capabilities.

Given this, our study uses GPT-4 as a representation of current LLM essay writing abilities. While in principle other LLMs could perform better, GPT-4's proficiency ensures that our test is a valid measure of the state of the art in AI-driven essay composition. 

\section{Method}\label{method}
\subsection{Overview}
This investigation entailed the blind assessment of 60 PDF documents, each comprising of five short-form essays, by five independent markers. This structure resulted in a collective evaluation of 1500 separate short-form essays with an average length of 285 words, combined into $n = 300$ graded submissions (60 PDFs $\times$ five markers) for evaluation purposes. The documents were equally divided between human-authored texts and those produced by OpenAI's GPT-4 model\footnote{Utilizing the gpt-4-1106-preview version accessed via OpenAI's API.}. These essays are components of the `Physics in Society' module facilitated by the Department of Physics at Durham University. The module's curriculum focuses on exploring the historical and philosophical dimensions of physics, including its evolution and the ethical implications surrounding the societal integration of its technological advancements. The module features a take-home essay assignment where students have 48 hours to respond to five questions in essays not exceeding 300 words each. Questions such as ``\emph{Is physics based on facts that follow from observations?}" and ``\emph{Is there a satisfactory interpretation of quantum mechanics?}" explore the history, philosophy, communication, and ethics of physics. The module's autumn semester provides formatively assessed questions to prepare students before they tackle a summatively assessed assignment on a new set of questions in the spring.

The study used three sets of student essays from 20 unique authors: the 2021/22 Formative assignment, the 2021/22 Summative assignment, and the 2022/23 Formative assignment. All questions from the assignments are shown in \ref{questionAppendix}. Although students submitted these essays as their own work, the presence of language models like GPT-2 since 2019 means authorship cannot be guaranteed. However, since ChatGPT was released on November 30, 2022, and mainstream adoption of AI chatbots began in 2023, essays submitted before this date are assumed to be student-authored. Essays from the 2022/23 Summative and 2023/24 Formative assignments were excluded from this study due to potential authorship ambiguity.

The evaluators assessed the essays based on their effectiveness in addressing five key elements, as outlined in Table \ref{markScheme}. Each submission, consisting of five essays, was marked holistically as a group, allowing students to strategically emphasize the aspects most relevant to the questions posed. For example, the question ``\emph{How did natural philosophers’ understanding of electricity change during the 18th and 19th centuries?}" encourages detailed discussions of specific physics concepts. This method mirrors the exact assessment process employed in the actual `Physics in Society' module at Durham University; all evaluators in this study had prior experience working within the module.

Grading followed the standard UK university criteria on a scale out of 100, with scores of 70\% and above qualifying for First-Class Honours, reflecting exceptional comprehension and skill. Next is Upper Second-Class Honours (2:1) for scores between 60\% and 69\%, considered a very good standard and often the minimum requirement for graduate positions and postgraduate study in the UK job market. This is followed by Lower Second-Class Honours (2:2) for 50\% to 59\%, Third-Class Honours (3rd) for 40\% to 49\%, and a Fail for marks below 40\%. Approximately 30\% of UK students achieve a First-Class Honours degree, a figure that varies by institution and subject but underscores the high standard of achievement these grades represent \cite{ofsDegree}. To ensure the integrity of degree classifications, UK universities typically require that marks do not vary significantly from year to year, resulting in overall averages that often hover around 65\%.

\begin{table}[htb]
\centering
\begin{tabular}{clp{12cm}}
\toprule
\textbf{Element} & & \textbf{Evaluation Criteria} \\
\midrule
1 & & Is there a high academic content, at a suitably advanced level, indicating familiarity with key milestones in the history of physics, the philosophy of physics, science communication, or ethics in academia? \\
2 & & Has the student formed an appreciation of the physics underlying a particular topic? \\
3 & & Does the student demonstrate a thorough grasp of the subject material? \\
4 & & How well does the student address the specific question asked? \\
5 & & Is the work written in a suitably authoritative, academic style, with material presented logically, coherently, and concisely, supported by appropriate factual information? \\
\bottomrule
\end{tabular}
\caption{Evaluation criteria for the essays, taken exactly from the `Physics is Society' module at Durham University.}
\label{markScheme}
\end{table}

Each element of the essays was graded on a scale from 0 to 100, in 5-point increments. Performance across these elements was categorized into seven distinct levels, and the average of these scores was calculated to give a final score out of 100. Scores from 80 to 100 indicated ``Exemplary" performance, showcasing exceptional insight and mastery beyond standard expectations. ``Excellent" scores ranged from 70 to 75, reflecting superior understanding and application, albeit not at the exemplary level. ``Good" (60-65) denoted solid competence and satisfactory execution, while ``Sound" (50-55) represented basic adequacy with some weaknesses. ``Acceptable" scores (40-45) signified marginal performance that just met minimum criteria, and ``Insufficient" (30-35) indicated notable deficiencies and a lack of basic understanding. The lowest category, ``Unacceptable" (0-25), marked failure to meet fundamental requirements, showing profound inadequacies in knowledge or execution. This grading framework provided a structured approach to evaluating essays, clearly delineating between varying levels of academic achievement.

In addition to marking the submissions for content quality and relevance, evaluators were tasked with assigning authorship to each essay using a four-point Likert scale. The options on the scale were `Definitely AI', `Probably AI', `Probably Human', and `Definitely Human'. This step of authorship assignment was deliberately conducted after the essays had been marked to prevent any potential bias in scoring based on the presumed origin of the text. The outcomes of these human evaluations are intended for comparison with results from various computational techniques designed for AI text detection. This comparative analysis aims to assess the efficacy of human judgement against automated methods in distinguishing between AI-generated and human-written texts.

\subsection{AI Essay Generation}
\label{ai-essay-gen}
The AI-generated essays were produced using OpenAI's API, leveraging the capabilities of the gpt-4-1106-preview model. Each question from the assignments underwent paraphrasing 10 times, incorporating a push for novelty. For instance, the question ``\emph{Was there a scientific revolution in 17th-century Europe?}" was transformed into ``\emph{In 250 words, analyze whether the 17th-century European developments, such as the Copernican model and Galileo’s telescope observations, truly signify a scientific revolution}". Similarly, ``\emph{Is there a satisfactory interpretation of quantum mechanics?}" was reworded to ``\emph{Examine in 242 words the de Broglie/Bohm theory as an alternative to mainstream quantum mechanics interpretations and its approach to wave-particle duality}". The specific suggestions for novelty in the prompts were sourced from the module syllabus. For example, the de Broglie/Bohm theory, the Copernican model, and Galileo's telescope observations are all covered as part of the course. This approach yielded 50 unique paraphrased prompts for each of the three assignments, resulting in a comprehensive collection of 150 prompts, detailed in the supplementary material\footnote{Also available at https://github.com/WillYeadon/AI-Exam-Completion}. The word count specified in the prompts often varied around 245, as this typically yielded essays close to 300 words in length.

The 150 AI-authored essays, generated from prompts, had a mean word count of 286.68 (SD = 26.44), comparable to the human-authored essays, which had a mean of 283.69 (SD = 13.51). A t-test indicated no significant difference in essay length between AI-generated and human-written essays (p-value = 0.219), suggesting that distinctions in content quality and complexity are the primary differentiators. Upon submitting the AI essays to Turnitin, the average plagiarism levels detected were 6\% for the 2021/22 Formative assessment, 0\% for the 2021/22 Summative assessment, and 1\% for the 2022/23 Formative assessment. This result shows that the AI-authored essays are novel.

\section{Analysis and Results}\label{section-analysis}
\subsection{Overview}
\label{subsec-analysis-overview}
\begin{figure}[!htbp]
\centering
\includegraphics[width=12.5cm]{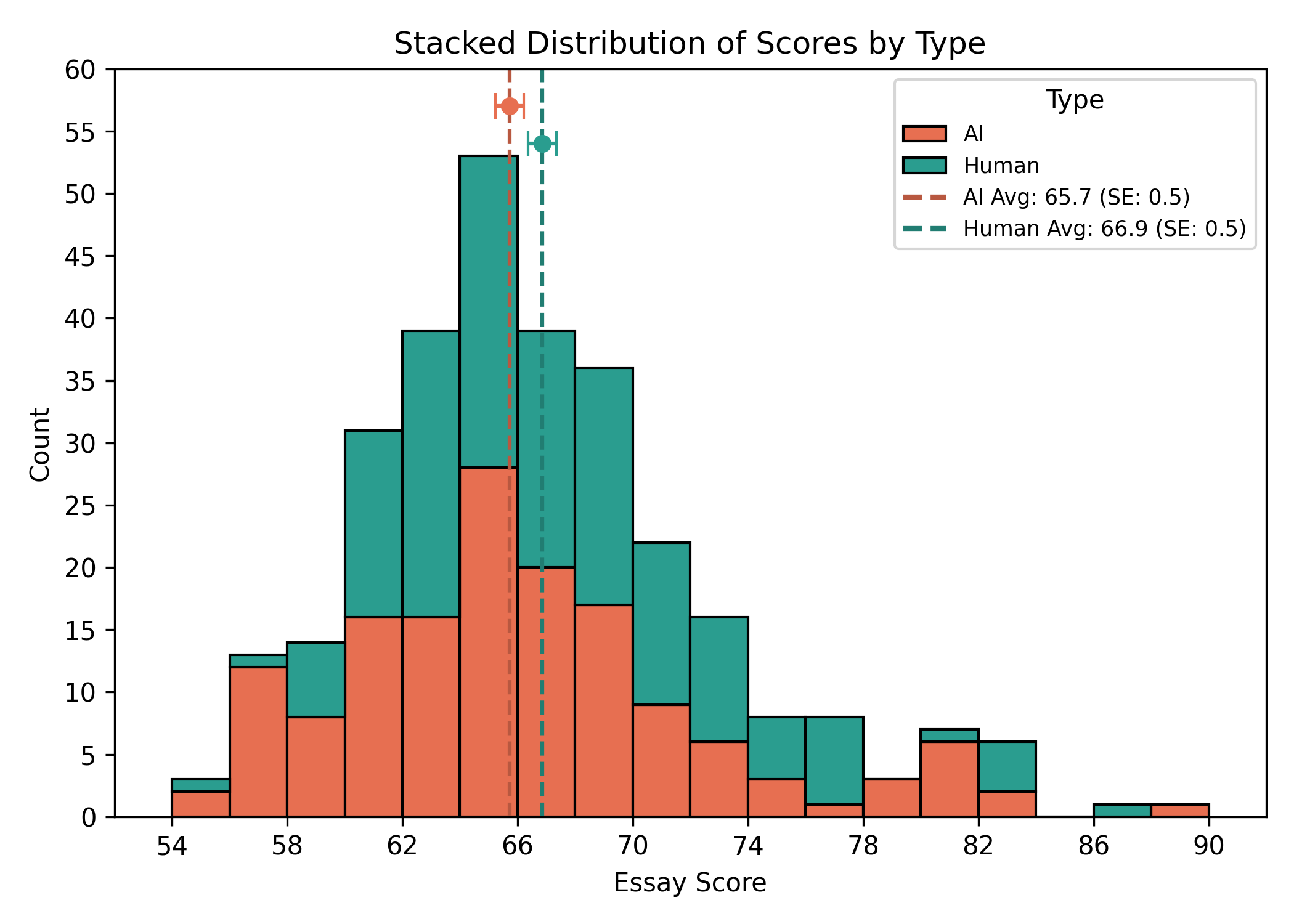}
\caption{Histograms of scores for AI and Human authored essays for all markers combined totalling $n = 300$ data points. The distributions look visually similar and a t-test ($\text{SD}_{\text{AI}} = 6.41$, $\text{SD}_{\text{H}} = 5.71$) reveals they are statistically indistinguishable.}
\label{fig-averages}
\end{figure}

The combined scores from five markers for sixty submissions resulted in a dataset of $n = 300$. Human-authored essays had an average score of 66.9 with a standard error of 0.5, while AI-authored essays averaged 65.7, with a standard error of 0.5. A t-test revealed a p-value of 0.107, indicating no statistically significant difference between the scores at an $\alpha$ level of 0.05. A histogram of the scores, as illustrated in Figure \ref{fig-averages}, shows similar distributions for both sets of essays. These findings suggest that AI and human authors are reaching parity in writing short-form physics essays, where assignments can be completed by ChatGPT within seconds and achieve scores comparable to those of human authors. For a detailed breakdown of the scores awarded by each marker and an analysis of the grading consistency among markers, please refer to \ref{byMarker}.

The decision to submit AI-generated work without knowing its potential score poses a risk, particularly for stronger students expecting high grades. In contrast, weaker students might find using AI advantageous, as the score distribution for AI-authored essays suggests that choosing a random essay from those created by the prompts detailed in Section \ref{ai-essay-gen} could result in a first or upper second-class grade 81.43\% of the time, according to the cumulative probability from a simple z-score ($\mu$ = 65.73, $\sigma$ = 6.41) calculation with a value of 60 - the boundary for a 2:1. These results echo those of Ghassemi et al.\cite{mollickCentaur}, who found that below-average performers gained more from using AI than above-average performers. The widespread availability of powerful language models is thus leveling the playing field in physics essay writing, offering students the opportunity to secure solid grades with minimal effort.

\begin{figure}[!htbp]
\centering
\includegraphics[width=12.5cm]{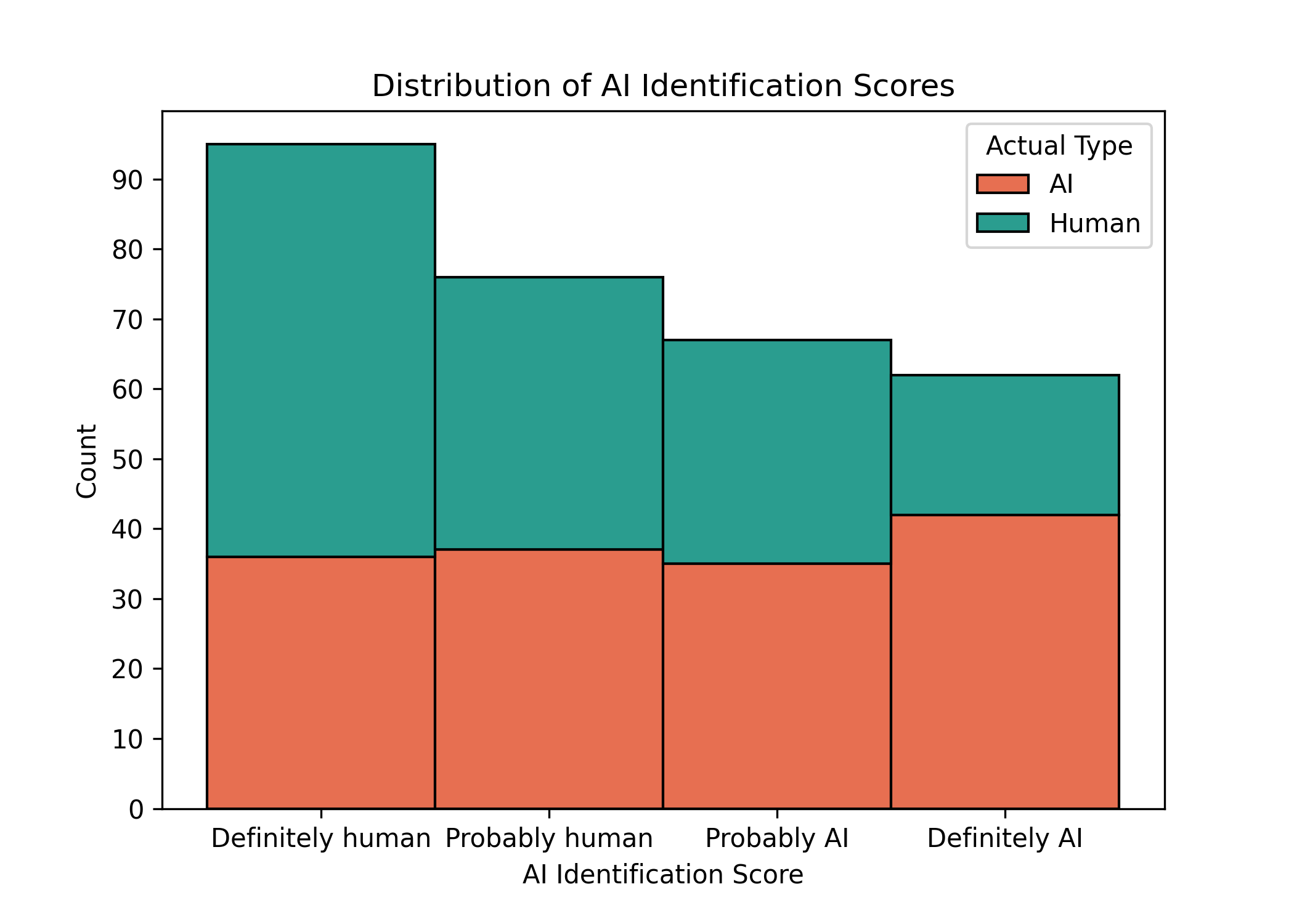}
\caption{Histogram showing the markers' assigned authorship versus actual authorship of the 300 assessed submissions. The amount of AI-authored texts is roughly constant in each category whereas the percentage of human-authored texts goes from 62\% 'Definitely human', 51\% in 'Probably human' then 48\% in 'Probably AI' and 33\% in 'Definitely AI'.}
\label{fig-ID-markers}
\end{figure}

After evaluating the essays, markers were asked to classify the authorship on a 4-point Likert scale, with 1 representing 'Definitely human' and 4 as 'Definitely AI'. The results, depicted in Figure \ref{fig-ID-markers}, showed a relatively constant amount of AI-authored texts across each category, while the proportion of human-authored text decreased from 'Definitely human' to 'Definitely AI'. This suggests a slight bias towards markers identifying essays as human-authored. Moreover, their confidence in the authorship was generally proportional to their accuracy. A Cochran–Armitage test for trend revealed a negative trend statistic (-0.093) for human-authored texts with a p-value of 0.027. This p-value is just below our $\alpha = 0.05$ indicating a mild at-best trend in markers being more confident in AI authorship as the proportion of human-authored texts decreases.

To better understand how markers can distinguish between essays written by AI and humans, we simplified their responses on the Likert scale to a straightforward yes or no decision, combining the `Definitely' and `Probably' categories. The accuracy rates recorded for the markers - 68\%, 67\%, 63\%, 57\% and 57\% - were only marginally better than a default 50\% success rate\footnote{Assuming a simplistic strategy where assigning all essays to a single category would result in 50\% accuracy, given an even split between AI and human-authored essays.}. Interestingly, markers were more likely to flag essays as AI-generated if they contained unusual features, such as the presence of em dashes (---), or if the essays were structured as a list of numbered points. Despite these indicators, the relatively low discernment success underscores the limitations of relying solely on human judgment to detect AI-generated work. Given the serious consequences of wrongly accusing a student of academic dishonesty, there is a pressing need for more reliable methods of authorship verification, such as employing lexical analysis or specialized AI detection tools.




\subsection{Lexical Analysis}
\begin{figure}[!htbp]
\centering
\includegraphics[width=15cm]{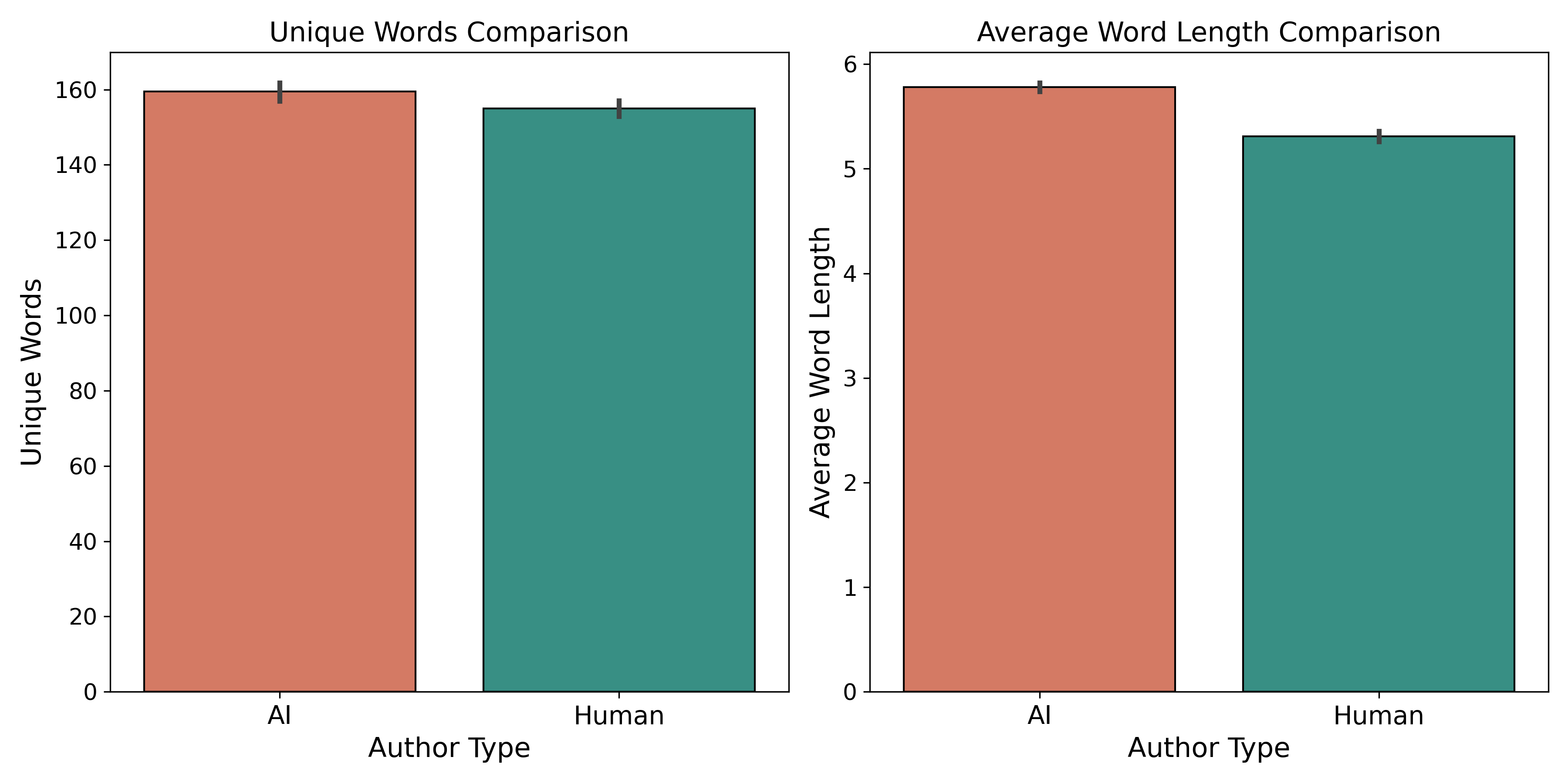}
\caption{Histograms comparing the distribution of unique words and average word lengths between human and AI-authored essays. There are statistical differences between the AI-authored and human-authored text for both metrics: unique words and average work length.}
\label{fig-lexical}
\end{figure}

An examination of the essays' lexical characteristics revealed that human-authored essays had an average of 154.96 unique words (SD = 12.77), while AI-authored essays featured 159.47 unique words on average (SD = 15.10). Additionally, the average word length was 5.31 (SD = 0.30) for human essays and 5.78 (SD = 0.24) for AI essays. Statistically significant differences were observed for both metrics, with p-values of $5.594 \times 10^{-3}$ for unique words and $3.454 \times 10^{-38}$ for average word length, indicating a slightly richer vocabulary in the AI-generated texts and a preference for longer words. The distribution and variance of unique words and average word lengths are visually depicted in Figure \ref{fig-lexical}. These findings suggest that AI essays not only employ a broader lexicon but also engage with complex language structures more frequently than their human counterparts. However, this lexical diversity and sophistication failed to translate into better scores; the mere presence of lexical richness does not guarantee profound comprehension or analytical insight. The true effectiveness of these essays is their ability to articulate a clear understanding of the underlying physics concepts and to address the posed questions with precision and depth.

\subsection{AI Authorship Computational Detection}\label{ai-detect-comp}
Following the introduction of ChatGPT, there was significant apprehension regarding its potential to automate exam responses, prompting investigations into its effectiveness on physics questions \cite{borChatGPT, yeadon2023exploring} and efforts to classify AI-generated texts \cite{pu2023deepfake, mitchell2023detectgpt}. Early on, it seemed that merely paraphrasing or editing AI-generated content could bypass existing detection technologies \cite{aiTextRel}. Some initial detection tools were criticized for their bias against non-native English speakers \cite{biasNonNative}. In response to these challenges, one of the leading plagiarism detection software companies, Turnitin, introduced an AI content detection feature. This move, however, was met with skepticism by numerous UK universities \cite{Staton2023} leading several institutions to forgo its use. Despite these historical challenges, advancements in AI content detection technology are steadily progressing, enhancing its accuracy and reliability.

\begin{figure}[!htbp]
\centering
\includegraphics[width=15cm]{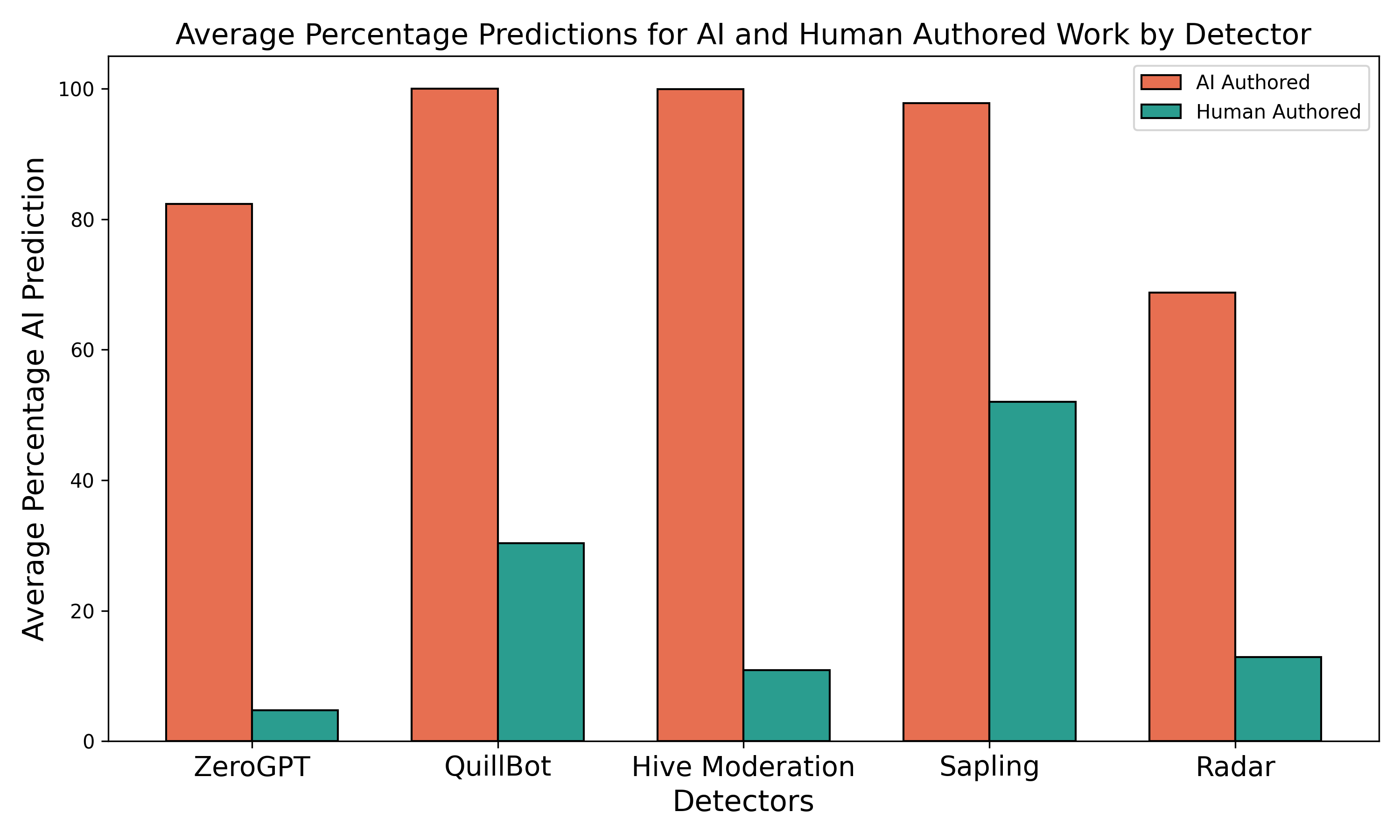}
\caption{Histogram showcasing the performance of five AI detection tools in differentiating between AI-authored and human-authored text. On average, all detectors rated the AI-authored content as more likely to be AI-generated than the human-authored text thou there was considerable variation in the percentages assigned to each submission.}
\label{fig-detector-comparison}
\end{figure}

To comprehensively examine the landscape of AI text detection, we evaluated the all the essays in our study using five distinct software tools: `ZeroGPT', `QuillBot', `Hive Moderation', `Sapling', and `Radar \cite{radar}'. These AI detection platforms employ varied methodologies to assess texts, yielding metrics such as Burstiness, Simplicity, Readability, and Perplexity. The output from these tools can range from categorical assessments (e.g., `Mostly AI-written' or `Partly AI-written') to a straightforward binary indication of AI content presence. Typically, these applications also quantify the likelihood of AI authorship in terms of a percentage confidence level. Figure \ref{fig-detector-comparison} offers a side-by-side comparison of the average percentages of the ``AI authorship" metric\footnote{The exact wording used varied depending on the detector.} assigned to AI-authored and human-authored submissions by each of the five detectors utilized in our study.

Applying a confusion matrix to the five AI text detectors, as shown in Table \ref{table:confusion_matrix_accuracy_precision}, reveals that `ZeroGPT' exhibits the best precision with no false positives (FP) and only one false negative (FN), achieving an accuracy of 98\%. Next, `QuillBot' shows a higher rate of false positives, mistaking human-written text for AI-authored content 10 times, which reduces its precision to 75\% and accuracy to 83\%. `Hive Moderation' records 4 FPs and no FNs, while `Sapling' has the highest number of FPs at 18. Compared to these proprietary models, `Radar' demonstrates moderate performance with a precision rate of 93\% and an accuracy of 88\%, although it has 5 FNs, suggesting that while it is quite reliable when it detects AI-generated content, it occasionally misses such content. Just as it is apt to measure generative AI quality with GPT-4 rather than GPT-3, the apparent state-of-the-art model, `ZeroGPT', shows very strong accuracy and precision in detecting AI text. 

This evaluation indicates that while AI-generated text may be detectable, the effectiveness of detection tools against content modified by paraphrasing remains uncertain. Our experiments with various paraphrasing tools suggest minimal impact on detection rates, which could reflect the limitations of these specific paraphrasing tools rather than an inherent robustness of detection algorithms. The possibility of human-assisted paraphrasing introduces an additional layer of complexity, effectively adding an `editorial phase' where AI-generated text is reviewed and modified by humans to evade detection at which point it may no longer be considered AI-text. Further, if paraphrasing significantly degrades the text's quality or coherence, its utility might be questioned. Thus, any integration of human judgment with AI-generated content ventures into a realm of ethical and practical ambiguity.

\begin{table}[ht]
\centering
\begin{tabular}{lcccccc}
\hline
Detector         & TP & FP & TN & FN & Accuracy & Precision \\ \hline
ZeroGPT          & 29 & 0  & 30 & 1  & 0.98     & 1.00      \\
QuillBot         & 30 & 10 & 20 & 0  & 0.83     & 0.75      \\
Hive Moderation  & 30 & 4  & 26 & 0  & 0.93     & 0.88      \\
Sapling          & 30 & 18 & 12 & 0  & 0.70     & 0.62      \\
Radar            & 25 & 2  & 28 & 5  & 0.88     & 0.93      \\ \hline
\end{tabular}
\caption{Confusion matrix components, accuracy, and precision for each detector. Of the detectors tested, `ZeroGPT' performs the best although the one open source detector, `Radar', shows good performance also.}
\label{table:confusion_matrix_accuracy_precision}
\end{table}

\section{Discussion}\label{section-disc}
\subsection{Impact on Higher Education}
The findings from this study compellingly indicate that short-form physics essays, when not invigilated, are highly susceptible to being completed by AI, rendering this assessment method ineffective. To safeguard the integrity of take-home essays, universities are faced with two options. One approach is to implement stringent measures to verify authorship, such as conducting oral examinations about the essays. Alternatively, institutions might place their trust in AI detection tools evaluated in this study, like ZeroGPT, to discern AI-generated content. Our analysis, conducted without specific prompt engineering or iterative AI improvement, revealed no discernable difference in quality between AI and human-authored essays, as unanimously determined by five independent markers. The imperative is clear: adapt take-home essay assignments or transition to in-person assessments.

A broader examination of the field of physics suggests that traditional physics exams with textural questions and answers largely resist completion by AI. GPT-4's performance on Durham University exams typically falls within a range of 40 to 50\% \cite{yeadon2023exploring} regardless of the academic level of the exam. Focusing on graphical interpretation, Polverini and Gregorcic \cite{BG-kinematics} found the performance from GPT-4 in kinematics graphs comparable to students who were taking a highschool-level course. However, the novelty of the responses led to some unusual errors not typically made by humans leading to different distributions in which questions were answered correctly. This echoed findings in \cite{yeadon2023exploring}, where GPT-4 approached physics problems using methods not covered in the syllabus. Further, both studies found that with multiple choice questions around 5\% of the time GPT-4 refuses to pick one of the available answers. Despite these challenges, there are opportunities within physics education as well. Correcting misconceptions often enhances educators' explanatory skills, suggesting that LLMs could aid teacher training by facilitating discussions on physics \cite{satisfiedSocrates}.  

While this investigation focused on physics essays, preliminary research in other academic areas hints at a likely convergence between AI and human authorship \cite{beowolf}. The code utilized in this study has been made available at \footnote{https://github.com/WillYeadon/AI-Exam-Completion} for educators wishing to conduct their own `AI Risk Assessment' in various disciplines. We encourage the use of this resource for evaluating the potential impact of AI on different fields of study.

\subsection{Limitations and Future Work}
The primary limitation of this study stems from the utilization of raw, unedited output from GPT-4 for the essays. Two significant considerations emerge from this approach. First, allowing the AI to iteratively refine its answers by identifying and amending its own mistakes could potentially enhance its knowledge in physics \cite{BG2}. Second, regarding assessment integrity, our analysis was limited to essays authored entirely by AI, rather than examining a blend of human and AI contributions. The use of 100\% AI-generated content could be seen as analogous to outsourcing essay writing to a paid service, a practice widely regarded as academic misconduct. However, the ethical landscape becomes murkier with minor AI contributions, such as using ChatGPT to rewrite a few sentences for improved clarity. The ambiguity increases further when considering tools like Microsoft Copilot, which are integrated directly into word processing software, presenting an evolving challenge without clear-cut solutions.

The findings of this study indicate no statistically significant differences between AI-generated and human-written essays. Consequently, conducting a randomized controlled trial where students are given varying levels of access to AI tools, and their essays are evaluated blindly, might not yield additional insights into the convergence of human and AI authorship quality. Given this context, future research should pivot towards understanding academic perspectives on the acceptable proportion of AI contribution in ``human-authored" work. The potential for computational identification of AI-text, highlighted in Section \ref{ai-detect-comp}, offers optimism for effectively detecting AI-generated content. The current paraphrasing tools' apparent limitations indicate that human intervention is required to render AI contributions undetectable. This situation presents an opportunity to establish national standards for the permissible amount of AI-generated content in academic documents. We propose that any work with $\leq 50\%$ AI-generated content be classified as human-authored as a starting practical guideline. This boundary is strictly weaker than requiring majority human-authored ($<50\%$ AI generated) content thus being more inclusive of a future where significant amounts of content will be created with AI assistance.

\section{Conclusion}
In the last 18 months, the rise of AI across various fields has been remarkable. Its impact necessitates considerable adjustments within physics education. This paper has shown that traditional non-invigilated short-form physics essays are losing relevance as humans and AI show similar writing abilities. This said, the outright dismissal of all non-invigilated exams may be premature; the burgeoning capabilities of AI detection technologies and idiosyncratic AI writing styles, like the inclusion of em dashes (---), suggest that it may be possible to limit egregious academic misconduct where AI work is copy/pasted and passed off as a student's own. Given this, we suggest $\leq 50\%$ AI-generated content to be the boundary for human-authored text as it is inclusive of a future with ubiquitous AI assistance whilst also respecting human-authorship.

This evolution is akin to the `hype cycle' model \cite{hypecycle}, where initial fervour for new technologies peaks and then wanes, only to stabilize as improvements in the underlying technology continue steadily. Consequently, educators must not panic but must evolve, incorporating AI into teaching and assessment strategies. This might involve assignments encouraging the exploratory use of tools like ChatGPT, allowing students to form their own opinions as to how they might utilize these powerful tools. While AI promises to augment the educational process, enriching rather than replacing the human touch, the authors would be reluctant to endorse a wholesale substitution of physics students with deep learning models \cite{replaceStudentsWithAI}!

\printbibliography

\newpage
\appendix
\section{Original Questions from ``Physics in Society"}\label{questionAppendix}
\begin{table}[htbp]
\centering
\begin{tabular}{clp{10cm}}
\toprule
\textbf{Question Number} & & \textbf{Question} \\
\midrule
1 & & Is physics based on facts that follow from observations? \\
2 & & What was the most important advance in natural philosophy between 1100 and 1400? \\
3 & & Was there a scientific revolution in 17th-century Europe? \\
4 & & How did natural philosophers' understanding of electricity change during the 18th and 19th centuries? \\
5 & & Does Kuhn or Popper give a more accurate description of physics? \\
\bottomrule
\end{tabular}
\caption{Questions used for the 2021/22 Formative Assignment.}
\end{table}

\begin{table}[htbp]
\centering
\begin{tabular}{clp{10cm}}
\toprule
\textbf{Question Number} & & \textbf{Question} \\
\midrule
1 & & If most theories have been shown to be false, do we have any reason to have confidence in our theories? \\
2 & & How has scientists’ understanding of charge changed during the 19th and 20th centuries? \\
3 & & Is there a satisfactory interpretation of quantum mechanics? \\
4 & & Was the project to build an atomic bomb typical of science in the twentieth century? \\
5 & & Why should the public believe scientists’ claims? \\
\bottomrule
\end{tabular}
\caption{Questions used for the 2021/22 Summative Assignment.}
\end{table}

\begin{table}[htbp]
\centering
\begin{tabular}{clp{10cm}}
\toprule
\textbf{Question Number} & & \textbf{Question} \\
\midrule
1 & & In your experience, is physics based on facts that follow from observations? \\
2 & & What, in your opinion, was the most important advance in natural philosophy between 1100 and 1400? \\
3 & & Was there a scientific revolution in 17th-century Europe? \\
4 & & How did natural philosophers' understanding of electricity change during the 18th and 19th centuries? \\
5 & & Would you judge that Kuhn or Popper gives a more accurate description of physics? \\
\bottomrule
\end{tabular}
\caption{Questions used for the 2022/23 Formative Assignment.}
\end{table}

\newpage
\section{Breakdown of marks by marker}\label{byMarker}
\begin{figure}[!htbp]
\centering
\includegraphics[width=15cm]{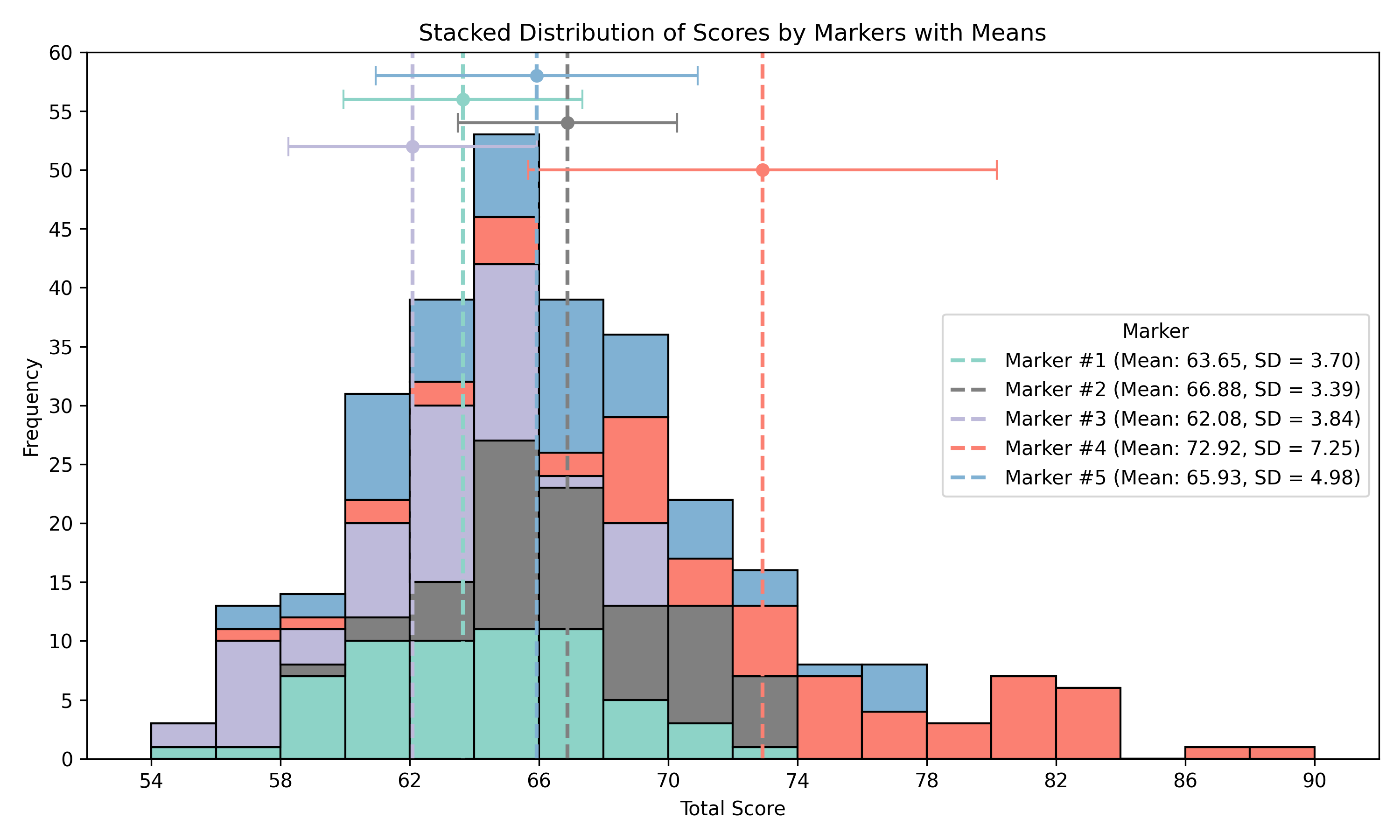}
\caption{Stacked histogram of the scores awarded by the five independent markers. Both the ANOVA and ICC models used find that the markers were not consistent in their evaluations.}
\label{fig-marker-stacked}
\end{figure}

A stacked histogram displaying the marks awarded by each marker is depicted in Figure \ref{fig-marker-stacked}. To assess the similarity of the marks assigned by the five markers, an Analysis of Variance (ANOVA) was performed, followed by Levene's test for equality of variances. The ANOVA revealed a highly significant p-value of $5.106 \times 10^{-29}$, indicating strong evidence against the null hypothesis that all group means are equal, suggesting that at least one marker's scores significantly differ from the others. Similarly, Levene's test yielded a significant p-value of $8.437 \times 10^{-10}$, indicating unequal variances among the groups, violating the assumption necessary for ANOVA.

To evaluate the consistency of grading among multiple markers for the 60 submissions, we employed the Intraclass Correlation Coefficient (ICC) as a statistical tool to examine grading reliability across markers. The ICC1 model was utilized to measure the absolute agreement among markers, with each submission being evaluated by a different set of markers selected randomly. This model specifically assesses the consistency of scores given to each submission. Conversely, the ICC2 model was applied to assess the extent of agreement among markers on the relative ranking of the submissions, rather than the exact scores. An ICC value of 1 indicates perfect agreement among markers, whereas a value of -1 signifies complete disagreement. The results from these models are presented in Table \ref{icc-table}.

\begin{table}[ht]
\centering
\caption{Intraclass Correlation Coefficient (ICC) Analysis Results}
\begin{tabular}{@{}lllllll@{}}
\toprule
Type & ICC & F & df1 & df2 & p-value & CI 95\% \\
\midrule
ICC1 & -0.053 & 0.749 & 59 & 240 & 0.907 & [-0.11, 0.03] \\
ICC2 & 0.035 & 1.323 & 59 & 236 & 0.076 & [-0.01, 0.11] \\
\bottomrule
\end{tabular}
\label{icc-table}
\end{table}

The results reveal a notable variability in grading consistency among markers in both ICC1 and ICC2 analyses, with ICC values nearing zero. This suggests that the grades for essays might have been significantly influenced by the markers' individual interpretations of the elements outlined in Table \ref{markScheme}. In the actual `Physics in Society' module at Durham University, only one marker is assigned per submission. However, given the observed variability among markers previously involved in the module, there is a potential risk of impact on students' academic outcomes unless efforts are made to align standards and minimize subjective variability.

Despite the observed variability in grading among markers, this study treats the evaluation as comprising $n = 300$ separate assessments, evenly divided between essays authored by humans and AI. This approach yielded an average score of 66.86 (SD = 5.70) for human-authored essays and 65.73 (SD = 6.41) for AI-authored essays. A t-test analysis revealed a p-value of 0.107, suggesting that there is no statistically significant difference between the scores of human and AI-authored essays at a significance level of $\alpha = 0.05$. This finding holds consistent across a more granular analysis, where the dataset is considered as five separate sets of $n = 60$ submissions, one for each marker. As detailed in Table \ref{t-test-by-marker}, this comparative analysis between human and AI-authored essays was replicated across all five markers, reinforcing the initial conclusion that there are no significant differences in scoring between the two groups.

\begin{table}[!htbp]
\centering
\begin{tabular}{lllll}
\hline
Marker     & Author & Mean      & Std       & p-value   \\ \hline
Marker \#1 & AI    & 63.07 & 4.15 &           \\
Marker \#1 & Human & 64.23 & 3.15 & 0.225   \\
Marker \#2 & AI    & 66.33 & 3.36 &           \\
Marker \#2 & Human & 67.43 & 3.39 & 0.212  \\
Marker \#3 & AI    & 61.37 & 4.47 &           \\
Marker \#3 & Human & 62.80 & 3.00 & 0.150  \\
Marker \#4 & AI    & 72.53 & 8.16 &           \\
Marker \#4 & Human & 73.30 & 6.34 & 0.686  \\
Marker \#5 & AI    & 65.33 & 4.53 &           \\
Marker \#5 & Human & 66.53 & 5.41 & 0.355  \\ \hline
\end{tabular}
\caption{Statistical results for t-tests for each marker individually showing that each marker's results also do not show a statistically significant difference in their means for AI-authored and human-authored essays.}
\label{t-test-by-marker}
\end{table}
\end{document}